\documentclass{icrc29}
\usepackage{graphicx,amssymb,amsmath,times}
\begin{document}
\title[TeV Gamma-Ray Spectra]{TeV
Gamma-Ray Spectra Unfolded for IR Absorption for a Sample of Low
and High Red-Shifted AGN}
\author[A. Konopelko et al.] {A. Konopelko$^{a,b}$, A. Mastichiadis$^c$, F.W.
Stecker$^d$ \\
        (a) Institute of Physics, Humboldt-University for Berlin, D-12489 Berlin, Germany\\
        (b) Max-Planck-Institute of Nuclear Physics, Postfach
        103980, Heidelberg, Germany\\
        (c) Department of Physics, University of Athens, Panepistimiopolis,
        GR 15783 Athens, Zografos, Greece\\
        (d) NASA Goddard Space Flight Center, Laboratory for High Energy
        Astrophysics, Code 662, Greenbelt, MD 20771
        }
\presenter{Presenter: A. Konopelko
(Alexander.Konopelko@mpi-hd.mpg.de), \
ger-konopelko-A-abs3-og23-poster}

\maketitle

\begin{abstract}
The H.E.S.S. collaboration recently performed the accurate
measurements of the $\gamma$-ray spectra above 200 GeV for two BL
Lac-type AGN in the Southern sky: PKS 2155-304 and PKS 2005-489.
The TeV spectrum of the BL Lac object 1ES 2344+514 has also been
recently measured by the Whipple collaboration. Using the results
of phenomenological calculations of the SED of the EBL we have
extracted the intrinsic $\gamma$-ray spectra of these blazars.
Consequently we have shown that these spectra can be fitted well
by means of a synchrotron self Compton model. \vspace*{-2mm}
\end{abstract}

\section{Introduction}

The ground-based very high energy (VHE) $\gamma$-ray astronomy,
exploiting imaging atmospheric \v{C}erenkov telescopes, continues
to enlarge the sample of BL Lac objects detected at energies above
100~GeV. Six of such objects, i.e. Mkn~421, Mkn~501, 1ES~2344+514,
1ES~1959+650, H~1426+428, and PKS~2155-304, have been confirmed in
observations with at least two instruments. The latter was
recently detected with the H.E.S.S. system of four imaging
atmospheric \v{C}erenkov telescopes in the Namib desert
\cite{hess-pks2155}. Another high-frequency peaked BL Lac
PKS~2005-489 was discovered by H.E.S.S. in the 2004 observational
campaign \cite{hess-pks2005}. The VHE $\gamma$-ray fluxes of these
sources are highly variable with a few clear indications of
variations of the spectral shape with the flux level, which seems
to be a generic feature of the objects of this class.

The measured TeV $\gamma$-ray spectra of extragalactic objects
could be strongly modulated due to absorption by interacting with
diffuse interstellar or intergalactic infra-red (IR) radiation.
Such attenuation caused by pair-production highly depends on
redshift of a TeV $\gamma$-ray source. Owing to the lack of direct
measurements of the IR background radiation in the wavelength
range 1-50~$\mu \rm m$, a computation of the opacity of the
intergalactic medium to TeV $\gamma$ rays remains model dependent.

In a simplified approach the most recent data on the extragalactic
background light (EBL) can be used to construct different
realizations representing all possible permutations between EBL
limits and the detections in the different wavelength regions
\cite{dwekkrennrich}. In general these realizations could help to
explore numerous possibilities of the intrinsic spectra of TeV
blazars. However, distinct absorbtion models (e.g.
\cite{malkanstecker,primack}) have inferred various observational
constrains and additional astrophysical information (e.g. galaxies
IMF), which finally mold a shape of the spectral energy
distribution (SED) of the EBL. Considering theoretically uncoerced
realizations of the SED of the EBL does not contribute much to the
absorption model and in particular to compiling a self-consistent
theory of emerging of TeV $\gamma$-rays out of blazar environment
and their propagation in the interstellar medium. Here we adopted
a different approach \cite{konopelko03}. We have folded the
distributions of the EBL derived in
\cite{malkanstecker,dejagerstecker} to construct the intrinsic
spectra using sets of TeV data for a number of blazars. These
spectra taken along with the contemporaneous X-ray data can be
fitted with a homogeneous synchro-self-Compton (SSC) model
\cite{mastikirk,konopelko03}. One can tightly constrain the
absorbtion model by testing it for an extended sample of TeV
blazars. Improved absorbtion model would reveal, in turn, an
appropriate information on the blazar intrinsic spectra.

\section{Observations}
\vspace*{-1mm}

The stereoscopic arrays of imaging atmospheric \v{C}erenkov
telescopes (IACTs), such as HEGRA and H.E.S.S., have achieved a
very good energy resolution of less than 15\% for spectral
measurements at energies above 100~GeV. Stereoscopic observations
allow also to tightly narrow down the systematic errors in the
measured energy spectra. Ultimately the accuracy of the spectral
measurement is determined by the statistics of the $\gamma$-ray
events above the cosmic ray background, which depends, in turn, on
source emission state and observational exposure.

{\bf Mkn~501 \& Mkn421.} The TeV energy spectra of two BL Lac
objects Mkn~501 and Mkn~421 have been studied in great detail with
the HEGRA system of five IACTs at La Palma, Canary Island. In 1997
Mkn~501 was in an unprecedented high state for almost 6 months. An
extremely high statistics of the accumulated TeV $\gamma$ rays
($>$30,000) provided very accurate measurement of the spectrum,
which reveals a gradual steepening towards the higher energies
\cite{ahar99}.  The energy spectrum of Mkn~421 was measured with
HEGRA in different states of emission at TeV energies up to
20~TeV. The energy spectrum of Mkn~421 is evidently curved at
higher energies, whereas data at energies below 3~TeV show
significant variations of the spectral slope depending on the flux
level \cite{ahar02}. The measured energy spectra of both Mkn~501
and Mkn~421 over the energy range from 500~GeV up to 20~TeV were
used to reconstruct the IR de-absorbed intrinsic source spectra,
which were then fitted along with contemporary X-ray data in a
homogeneous SSC model \cite{konopelko03}.  In \cite{konopelko03} a
logically consistent model of X-ray and TeV $\gamma$-ray emission
was constructed for both AGN, Mkn~501 and Mkn~421, taking into
account IR absorption according to the SED of the EBL as given in
\cite{malkanstecker,dejagerstecker}.

{\bf PKS~2155-304.} PKS~2155-304 has been observed with a single
telescope and the system of two H.E.S.S. telescopes, during
construction phase of the experiment, in a few exposures during
year 2002 and 2003 for a total observational time of about 52~hrs
\cite{hess-pks2155}. The overall significance of the $\gamma$-ray
signal is of 45$\sigma$ and about 4,500 excess events registered
above 160~GeV. This unique data sample of PKS~2155-304 provided
very accurate measurement of the TeV energy spectrum. The energy
resolution for each individual $\gamma$ event was better than
15\%. The data clearly indicated high variability of PKS~2155-304
in TeV $\gamma$-rays between 10\% and 60\% of constant Crab Nebula
flux, but no evidence found that the spectral index varies with
time. It justified the measurement of the time-averaged spectrum,
which is rather steep in the energy range from 200~GeV up to
2.5~TeV. The spectrum can be well fitted by pure power-law with a
spectral index of $3.32\pm0.06$. Despite that the residuals at
high energies are in fact smaller for a power-law fit with an
exponential cutoff at $\simeq$1.4~TeV, this PKS~2155-304 data does
not reveal a statistically firm indication of the exponential
cutoff at TeV energies. The shortest variability time scale was
found to be of half an hour. We used here this time-average
spectrum of PKS~2155-304 measured by H.E.S.S. \cite{hess-pks2155}
to extract the intrinsic emission spectrum by unfolding the IR
absorption. The steep TeV energy spectrum of PKS~2155-304 caused
by IR absorbtion was predicted in \cite{fs}.

{\bf PKS~2005-489.} Recently the $\gamma$ rays above 200~GeV with
the 2.5\% flux of that from the Crab Nebula has been detected from
another high-frequency peaked BL Lac PKS~2005-489 using the full
H.E.S.S. array of IACTs in the 2004 observational campaign for an
exposure of 24.2~hrs \cite{hess-pks2005}. The statistical
significance of the signal was of 6.7$\sigma$ with about 300
excess events. The energy spectrum is consistent with a pure
power-law with an exponent of $4.0\pm0.4$. These data do not allow
conclusions on any sophisticated spectral shape as well as on the
time variations of the spectral slope. Non-detection of
PKS~2005-489 with H.E.S.S in 2003 observational campaign
ascertained a clear variability of the source in TeV $\gamma$
rays, even though this data can not be effectively used to
establish a shortest variability time scale. The spectral data
points given in the energy range expanding up to 2.3~TeV
\cite{hess-pks2005} are characterized by rather large statistical
errors at high energies, which can be naturally explained by the
low statistics of $\gamma$-ray events in the corresponding energy
bins. We have used this data for the consistency check of the IR
absorption model.

{\bf 1ES~2344+514.} The BL Lac object 1ES~2344+514 is an
established source of TeV $\gamma$-rays (see \cite{1es2344} and
references herein). The brightest outburst in TeV $\gamma$ rays
from this object was detected with a 10~m VERITAS telescope on the
night of December 20, 1995. During this flare the $\gamma$-ray
flux exceeded by about a factor of two the flux of the Crab
Nebula, which provided the detection of 128 excess events at the
confidence level of 5.3$\sigma$. So far this is the best data
sample available for the spectrum evaluation of the 1ES~2344+514.
The resulting spectrum between 0.8~TeV and 12.6~TeV can be
described by a power-law of spectral index of $2.54\pm0.17$, with
given error as the statistical one. For the energy resolution of
VERITAS telescope at Mt. Hopkins of about 40\% and very small
signal, rather coarse bins are preferable for the spectral data
presentation (four bins per decade). Detailed studies of spectral
evaluation procedure allowed to substantially limit all possible
systematic energy biases \cite{1es2344}. Note that both
1ES~2344+514 and 1ES~1959+650 are located at almost the same
redshift. Taken together they form a second pair of TeV blazars at
the same redshift after a famous pair of Mkn~501 and Mkn~421.

\begin{table}[t]
\caption{\label{bllacs} Summary of the measured TeV
$\gamma$-spectra $(dN/dE)_{m}$ for a concurrent sample of seven BL
Lac objects. $\nu_{TeV}$ corresponds to the estimated IC peak
position for the IR de-absorbed spectra.} \vspace*{1mm} \centering
\begin{tabular}{||l|l|l|l|l||} \hline \hline
~Object~ & ~Redshift~ & ~$(dN/dE)_{m}$~ & Ref.: & $log \,\, \nu_{TeV}$ [Hz] \\
\hline \hline
Mkn~421 & 0.031 & $\propto E^{-2.19}e^{-E/3.6\,\, TeV}$ & \cite{ahar02} &  $\simeq$26.5 \\
Mkn~501 & 0.034 & $\propto E^{-1.92}e^{-E/6.2\,\, TeV} $ & \cite{ahar99} & $\simeq$27.5 \\
1ES~2344+514 & 0.044 & $\propto E^{-2.54}$ & \cite{1es2344} & $\simeq$26\\
1ES~1959+650 & 0.048 & $\propto E^{-2.83}$ & \cite{1es1959} & $\simeq$26\\
PKS~2005-489 & 0.071 & $\propto E^{-4.0}$ & \cite{hess-pks2005} & $\leq$25.5 \\
PKS~2155-304 & 0.117 & $\propto E^{-3.32}$ & \cite{hess-pks2155} & $\leq$25.5 \\
H~1426+428 & 0.129 & $\propto E^{-3.5}$ & \cite{petry} & $\leq$25.5\\
\hline
\end{tabular}
\vspace*{-2mm}
\end{table}

{\bf 1ES~1959+650.} The BL Lac 1ES~1959+650 was detected with the
HEGRA system of IACTs in a high state during May-June 2002 with
the flux at TeV energies exceeding of that from the Crab Nebula
\cite{1es1959}. The 8.5~hrs exposure provided detection of 255
$\gamma$ rays at the significance level of 25.5$\sigma$. In the
energy range between 1.5~Tev and 10~TeV the data were fit to a
power law, which results in the spectral index of $2.83\pm 0.22$.
At the same time the power law fit of spectral index of $1.83\pm
0.23$ with the exponential cutoff of about 4.2~TeV describes
equally well the data. For given statistical and systematic errors
of the power-law spectral index the energy spectra of 1ES~1959+650
measured with HEGRA in high and low states are marginally
undistinguishable.


{\bf H~1426+428.} The H~1426+428 has the largest redshift,
z=0.129, among all of the so far established TeV blazars. The
energy spectrum of H~1426+428 as measured by VERITAS collaboration
\cite{petry} is a power law with a spectral index of $3.5\pm 0.4$
in the energy range from 250~GeV up to 2.5~TeV. Such steep energy
spectrum is consistent with the spectrum for another BL Lac object
PKS~2155-304 (see above) located at similar redshift of z=0.117.
However, observations of H~1426+428 with the HEGRA system,
characterized by rather low $\gamma$ rays statistics of about 200
events and signal significance of about 6$\sigma$, revealed very
peculiar spectral shape above 1 TeV \cite{horns}, which
constitutes for a statistically significant deviation of the
spectrum from a power law. It is apparent that further
observations of this source are needed in order to reexamine its
TeV energy spectrum. We have omitted discussion of these data in
this paper.

\begin{figure}[t]
\centering
\includegraphics*[width=0.55\textwidth,angle=0,clip]{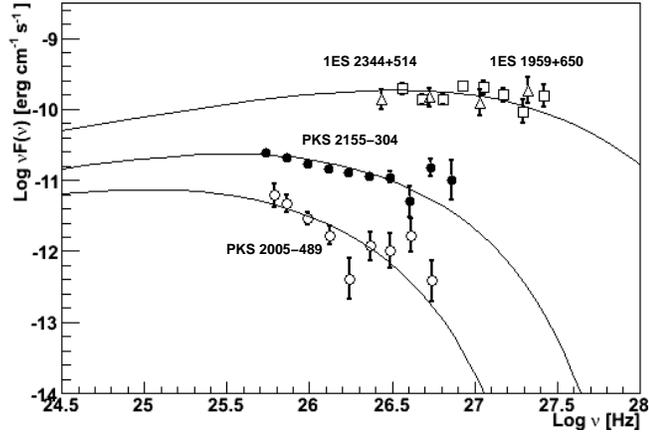}
\vspace*{-2mm} \caption{\label {fig1} The de-absorbed TeV
$\gamma$-ray spectra of four BL Lac type objects: PKS~2005-489,
PKS~2155-304, 1ES~2344+514 ($\bigtriangleup$), 1ES~1959+650
($\Box$). The IC spectra computed in the homogeneous SSC model are
also shown. }
\vspace*{-2mm}
\end{figure}

\vspace*{-3mm}
\section{Discussion on the IR de-absorbed spectra}
\vspace*{-1mm}

We have used here the empirically based model for the EBL
developed in \cite{malkanstecker,dejagerstecker}. This model is
consistent with all available data on the EBL over the 1-300~$\mu
\rm m$ wavelength range taking into account the statistical and
systematic errors of those measurements. The SED of the EBL was
used to compute the opacity of TeV $\gamma$-rays, $\tau(E,z)$, as
described in \cite{konopelko03}. Finally we have reconstructed the
intrinsic source spectrum using the measured spectrum as
$(dN_\gamma/dE)_i = (dN_\gamma/dE)_m \cdot exp[\tau(E,z)]$. The IR
de-absorbed spectra are shown in Figure~\ref{fig1}.  We do not
observe any non-physical features, e.g. sharp upturns etc, in any
of the de-absorbed spectra. All spectra are well consistent with a
generic IC TeV spectrum. Thus the X-ray and TeV $\gamma$-ray data
for PKS~2155-304, which is the most distant BL Lac from those
discussed here, can be well fitted in the homogeneous SSC model
using parameters: $\rm B=0.5\, G$, $\delta=50$, and
$\gamma_{max}=2\times10^5$. Such values of the model parameters
are very close to the corresponding values obtained for Mkn~421
and Mkn~501 \cite{konopelko03}. We note that there is an apparent
trend in steepening of the measured TeV spectra with the
enlargement of the redshift. We plan to examine this more closely
in a forthcoming paper. However, the preliminary results presented
here show that the IR-absorption/SSC model developed in
\cite{konopelko03} for two low redshift BL Lac objects, Mkn~421
and Mkn~501, can be applied as well to the new set of
extragalactic TeV sources.

\end{document}